# Explainable deep learning for insights in El Niño and river flows

Yumin Liu[1,2], Kate Duffy[3,4,5], Jennifer G. Dy[1,2], and Auroop R. Ganguly[2,3,,6]*

1. SPIRAL Center, Department of Electrical and Computer Engineering, Northeastern University, Boston, Massachusetts 02115, USA

2. The Institute for Experiential AI (EAI), Northeastern University, Boston, MA, USA

3. Sustainability and Data Sciences Laboratory, Department of Civil and Environmental Engineering, Northeastern University, Boston, Massachusetts 02115, USA

4. NASA Ames Research Center, Moffett Field, California 94035, USA

5. Bay Area Environmental Research Institute, Moffett Field, California 94035, USA

6. Pacific Northwest National Laboratory, Richland, WA 99354, USA

* Corresponding author

**The El Niño Southern Oscillation (ENSO) is a semi-periodic fluctuation in sea surface temperature (SST) over the tropical central and eastern Pacific Ocean that influences interannual variability in regional hydrology across the world through long-range dependence or teleconnections. Recent research has demonstrated the value of Deep Learning (DL) methods for improving ENSO prediction as well as Complex Networks (CN) for understanding teleconnections. However, gaps in predictive understanding of ENSO-driven river flows include the black box nature of DL, the use of simple ENSO indices to describe a complex phenomenon and translating DL-based ENSO predictions to river flow predictions. Here we show that eXplainable DL (XDL) methods, based on saliency maps, can extract interpretable predictive information contained in global SST and discover SST information regions and dependence structures relevant for river flows which, in tandem with climate network constructions, enable improved predictive understanding. Our results reveal additional information content in global SST beyond ENSO indices, develop new understanding of how SSTs influence river flows, and generate improved river flow predictions with uncertainties. Observations, reanalysis data, and earth system model simulations are used to demonstrate the value of the XDL-CN based methods for future interannual and decadal scale climate projections.**

**Introduction**

The El Niño-Southern Oscillation (ENSO) is a primary mode of interannual weather variability around the globe. ENSO modulates flood timings in Africa[1], interannual variability of flow in the Ganges, the Amazon, and the Congo rivers[2,3], and has significant influences on regional climate and hydrologic patterns around the globe. A predictive understanding of ENSO is thus of economic and societal importance. However, and our ability to predict ENSO with physics-based numerical simulations or data-driven models at interannual, decadal, and multidecadal time horizons have remained relatively poor[4], which has in turn hindered our ability to assess and leverage the predictability of ENSO's hydrometeorological effects.

Some challenges in ENSO forecasting may be traced back to data limitations, such as the relatively arbitrary rectangular regions that determine ENSO indices. Studies have suggested that ENSO is part of a larger system of interrelated SST oscillations which may co-impact regional hydrometeorology[5]. Further, our understanding of physical mechanisms[6] along with data-driven methods[7] suggest that the relationships between ENSO and river flows may be highly nonlinear. The resulting complexity of the earth system calls for methods that can leverage complete information content from global SST data and identify complex geographic dependence structures, which include both proximity-based dependence and long-range teleconnections. Fig. 1 shows SST anomalies in year 2008 when there was a cool year (La Nina phenomenon), while Fig. S1(a) and S1(b) show SST anomalies in a warm year (El Niño) and a neutral year, respectively. The relationships between river flows and ENSO indices indicate the possibility of significant nonlinear dependency (Table S3 and Fig. S2 and S3).

Commonly used methods to identify dependencies among climate variables include visual comparison[8], correlation[9], mutual information[7], coefficient of determination[10], and weights in (sparse) linear regression[11,12]. These methods often require heuristic expertise in selecting features and can be difficult to extend to more complex features such as three-dimensional spatiotemporal features. In the recent years, deep learning methods have seen preliminary success in climate science, meteorology, and hydrology, resulting in improved predictive skills and development of new methods to investigate the spatiotemporal dependencies[13,14]. Furthermore, methods for interpretation and explanation of deep neural networks, such as saliency maps, can be adapted to climate problems to analyze relevant (SST) regions resulting in understandable predictive information for regional climate and hydrology. Simonyan et al.[15] initially proposed the saliency map method as a visualization technique to explain the neural network function mapping, specifically, the extent to which inputs contribute to network output. Due to their effectiveness, explainable deep learning methods have been widely applied to the geosciences and especially to understand climate science and translate to impacts, for example, in spatial drought prediction[16], satellite-based PM2.5 (air pollution) measurements[17], crop yields[18], species distribution models[19], analysis of hailstorms[20], hydro-climatological process modeling[21], precipitation quality control[22] and climate drivers for global temperature[23], and to localize pest insects in agricultural application[24]. Ham et al.[13] used saliency map to analyze which regions contributed most in predicting the Niño 3.4 index using their neural network. Similarly, Mahesh et al. [25] applied saliency maps to find the important geographic regions for predicting Niño 3.4 index.

Here we address the problem of developing explainable predictive insights relating to the ENSO phenomenon. Our approach is based on an eXplainable Deep Learning (XDL) solution [15] that concurrently uses convolutional neural networks (CNN) for prediction of river flow time series and saliency maps to explain the results by highlighting the relative importance of the spatiotemporal SST data. Our implicit hypothesis is that the XDL approach will lead to advances in predictive skills of river flows by considering the information content in the entire SST map, which should exceed the information content of ENSO indices. Furthermore, the XDL approach may lead to new discoveries of robust SST teleconnections with each other and with river flows, which in turn would further explain the gains in predictive skills. We develop correlation-based metrics to quantify SST autocorrelations and teleconnections either owing to known proximity-based spatial correlations or owing to known long-range spatial dependence. The approaches are developed for proxy observations (reanalysis) datasets as well as earth system model (ESM) simulated Coupled Modeling Intercomparison Project phase 5 (CMIP5) data, both for assessments of historical skills as well as for use in future projections of teleconnections and river flows which represent a major gap in current generation earth system models[26–28].

**Results and Discussion**

We trained a CNN (Fig. S4) to predict monthly Amazon and Congo River flow from monthly SST derived from Earth System Models (ESM) and reanalysis data. We compared the skill to that of an ensemble of ML models, which predicted river flow using only indices calculated from the Niño 3.4 region (5°S-5°N, 170°W-120°W). These indices include mean SST over the Niño 3.4 region as observed and modeled in reanalysis and ESMs, as well as the Niño 3.4 index, an anomaly value. We found that models with access to the larger SST area (41.5°S-37.5°N, 50.5°E-9.5°W), with its full spatial and temporal provenance. outperformed models using the ENSO indices for prediction of three-month rolling mean river flows on both the Amazon on the Congo River (Fig. 2). The CNN ingesting more SST information also outperformed the historical climatological mean as a predictor of the Amazon and Congo River flows. This suggests the larger SST region was useful for capturing the phase and amplitude of annual river flow fluctuations as well as components of interannual variation. Predictive information on the interannual variability of the Amazon River flow was either not fully expressed in the ENSO region, or else was not captured by the ensemble of ML models (linear regression, lasso regression, ridge regression, elastic net regression, random forest regression, and feed forward dense neural network, or DNN, regression).

For SST as a predictor of river flow, seasonality was not removed to avoid potential information loss when delineating between anomaly and climatological states, which may be imprecise due random-frequency climate variability with periods exceeding typical climatological timescales. Thus, the task of the SST models was to predict the temporal climatology of river flow values. With the Niño 3.4 index as a predictor of river flow anomaly, and seasonality was subsequently added back to the river flow value. For the Amazon River, we found that all models using climatological SST in the Niño 3.4 outperformed models using the Niño 3.4 anomaly.

The task of predicting Congo River flow was more challenging, perhaps influenced by the more extensive management of the Congo River basin compared to the Amazon River basin. However, predictions based on reanalysis model SST still resulted in lower RMSE than baseline predictions based on historical climatological mean for the Congo River. In most cases, Congo River flow predictions based on Niño 3.4 anomaly value (index) outperformed predictions based on the climatological value of SST in the Niño 3.4 region. A full comparison of RMSE for river flow prediction using indices and larger area SST is presented in Table S2.

Prediction of river flow using zero lag (concurrent) SST data is relevant to predicting future river flow in climate projections. Mappings between observations of river flow can also give insight into the predictability of the system; deeper analysis of CNN performance and historical average (presented in Tables S4 and S5) suggest that the methods compare differently when different aspects of performance (linear/nonlinear correlation, seasonal/yearly, extremes, etc.) are examined. For example, the ESM+CNN model achieved a lower mean absolute error and stronger association with Amazon River flow by metrics of linear correlation than the climatological mean, but has a higher RMSE in spring, when Amazon River discharge generally peaks.

We used a cyclical saliency map method to identify important spatial areas for the network to make predictions of river flows (Fig. 3). From the saliency maps we discover that the predictive power of ESMs comes mainly from the ENSO and the Indian Ocean Dipole (IOD) regions, suggesting a strong link between these two phenomena and a co-impact on regional hydrology. Fig. 3(a) shows that the dominant salient areas for Amazon River flow prediction are in tropical Pacific and Indian Oceans. Fig. 3(c) shows similar patterns but with less strong and smaller salient areas for Congo River flow. When using reanalysis data (Fig. 3(b) and 3(d)), the saliency maps are much more diffused, suggesting that the CNN model does not pick up any strong relationships between the predictor and predictand. However, the presence of linear and nonlinear information content about river flow in global SST is confirmed by the maps in Fig. S6. The yearly cyclical saliency maps and seasonal saliency maps are also presented in the Fig. S5-S8. Whereas saliency maps can be used to verify the physically reasonable relationships that are learned as well as to discover new relationships, our hypothesis can be confirmed by examining the degree to which known oceanic regions that correspond to the ENSO region, as well as oceanic regions that correlate with the ENSO region, are triggered by the saliency maps as contributors to the information content.

Complex network theory provides a complementary tool to investigate the short and long-distance relationships in earth systems, such as teleconnections associated with the ENSO phenomenon that are indicated by our results. We analyzed the correlation structure of global SST data by constructing degree maps for reanalysis and ESM SST (Fig. 4). We quantified temporal correlation by calculating Pearson's correlation coefficient between every pair of locations in the ocean. The degree of each geographical location is the number of edges connected to this location, where an edge exists if the correlation is larger than a threshold $c_1$. We also set a second correlation threshold $c_2$ and distance threshold $d$ to define a teleconnection. We define that there is a teleconnection between two locations if their distance is larger than $d$ km and the correlation is larger than $c_2$.

We find that ESM SST has high degree values over a large area, indicating that the SST are highly correlated through both proximity-based correlations and teleconnections. There are many teleconnections between tropical Pacific Ocean, Indian Ocean, and even Atlantic Ocean, and they are largely concentrated around the equator (Fig. 4(a)). The teleconnections remain strong when the correlation threshold is increased (Fig. 4(c)). This pattern is reflected in the histogram of edges, which shows the degree distribution (Fig. 4(e) and 4(g)). There are many edge counts for long distances, which demonstrate the multicollinearity between SST regions. In contrast, the histograms of edges for reanalysis data (Fig. 4(b) and 4(d)) show fewer long-distance connections for a low correlation threshold, and negligible long-distance connections with a high correlation threshold. These results indicate a weaker correlation structure in reanalysis SST compared to ESM SST, and are consist with recent literature indicating that ESMs tend to exhibit a stronger coupling than reanalysis or observations[29–32]. Extending these findings, a hypothesis for future studies by climate science and earth system modeling communities is that the coupling strength of ESM model components are usually stronger than those in observations or reanalysis, and that data-driven sciences may be able to quantify and bridge this gap.

Histograms of connection distance in each of ESMs indicate qualitative differences in the correlation structures of the models (Fig. S9 and S10); some exhibit a single peak corresponding to proximity-based correlations (e.g. Fig. S9(a)), while others also exhibit clusters of long-range connections (e.g. Fig. S9(f)). Models also vary in the rapidity of decay of proximity-based correlations with increasing distance. These attributes of these plots indicate distinct spatiotemporal correlation structures among the climate models.

ENSO is a complex spatiotemporal process with global impacts on SST and the flows of large rivers globally, especially around the tropics and subtropics. In this work we combined ML methods and interpretive techniques to obtain gains in predictive power and make new discoveries about dependence structures and teleconnections in global SST data. Although researchers often analyze the relationship between ENSO indices and the other climate variables, our results indicate that information outside of the canonical ENSO region can help to predict regional hydrology better than some representations based on hand-selected features. They suggest that additional data and data-driven technologies could lead to a better understanding of mechanisms and the flow of causality in earth systems, as well as to informed climate adaptation through augmented projections of river flow for future climate scenarios.

**Methods**

Flowcharts detailing the methodology are provided in Fig. S11. The processing, modeling, and evaluation steps are outlined for reanalysis data (Fig. S11(a)) and ESM data (Fig. S11(b)). The ensembling approach used to generate probabilistic river flow predictions is shown in Fig. S11(c).

**Datasets**

We obtained monthly sea surface temperature datasets from ESM simulations and reanalysis models. The ESM datasets are downloaded from NASA Earth eXchange (NEX, https://registry.opendata.aws/nasanex/, last access May 2021). From the full set of Coupled Model Intercomparison Project Phase 5 (CMIP5) ESMs by various institutes, we discard those which have some months missing, leaving 32 ESMs. The CMIP5 historical forcing experiment spans from January 1950 to December 2005, or 672 months in total. This ESM dataset covers the whole globe with a spatial resolution of 1° longitude by 1° latitude (approximate 100km by 100km) with longitudes range from 0.5°E to 359.5°E, and latitudes from 87.5°N to 87.5°S. The ESM names are shown in Table S1.

In addition to ESM simulation datasets, we also use reanalysis datasets which are combinations of sparse on-site observation with other sources (such as remote sensing and satellite imaging) to produce gridded data. It is common to use reanalysis data as the proxy of true observational data because the site-based observational data are very sparse and not gridded. We use three reanalysis datasets in the experiment as predictors: Hadley-OI SST dataset[33], COBE SST dataset[34] and ERSSTV5 dataset[35].

The merged Hadley-OI SST dataset (https://climatedataguide.ucar.edu/climate-data/merged-hadley-noaaoi-sea-surface-temperature-sea-ice-concentration-hurrell-et-al-2008) is a combination of two reanalysis datasets: HadISST1[36] and NOAA OI.v2[37]. The HadISST1 dataset is derived gridded, bias-adjusted in situ observations, and the NOAA OI.v2 dataset combines in situ and satellite-derived SST data. The resulting Hadley-NOAA-OI dataset contains monthly mean sea surface temperature from year 1870 to 2020 with a spatial resolution of 1° longitude by 1° latitude.

The COBE SST dataset (https://climatedataguide.ucar.edu/climate-data/sst-data-cobe-centennial-situ-observation-based-estimates) are centennial in situ observation-based estimation that combines SSTs from International Comprehensive Ocean-Atmosphere Data Set (ICOADS)[38] release 2.0, the Japanese Kobe collection and reports from ships and buoys. ICOADS is the most comprehensive archive of global marine surface climate observations available, but the data coverage is sparse and neither gridded nor corrected. These datasets were gridded using optimal interpolation. The resulting COBE dataset contains monthly mean sea surface temperature from 1891 to 2020 with a spatial resolution of 1° longitude by 1° latitude.

The NOAA extended reconstruction SSTs version 5 (ERSSTV5) dataset (https://climatedataguide.ucar.edu/climate-data/sst-data-noaa-extended-reconstruction-ssts-version-5-ersstv5) is based on statistical interpolation of the ICOADS release 3.0 data and Argo (https://argo.ucsd.edu/) float data. The resulting ERSSTV5 dataset contains monthly mean sea surface temperature from year 1854 to 2019 with a spatial resolution of 2° longitude by 2° latitude.

These datasets have different time spans and spatial resolutions. We performed preprocessing to align the coordinates, interpolate to the same spatial resolution by bilinear interpolation, and select the common time span. A minimal number of missing values were filled with 0, in a similar approach to the zero padding approach in machine learning, where a matrix is surrounded with zeroes to help preserve features at the image edges. After preprocessing, the resulting reanalysis input has 3 channels corresponding to the 3 reanalysis datasets described above with a spatial resolution of 1° longitude by 1° latitude. We extract the region with latitude from 37.5°N to 42.5°S and longitude from 50.5°E to 0.5°W, roughly covering most of low latitude Pacific Ocean and Indian Ocean. The resulting input image size is 80×300 height by width.

The Niño 3.4 SST Index time series is anomaly monthly average SST in the region with latitude from 5°S to 5°N and longitude from 170°W to 120°W with the 1981-2010 mean removed. The data is generated by the NOAA Physical Sciences Laboratory using the HadISST1 dataset[36].

The river flow dataset was obtained from UCAR (A. Dai 2017) and can be downloaded from UCAR Research Data Archive website (https://rda.ucar.edu/datasets/ds551.0/index.html, last accessed January 2021). The dataset contains monthly runoff ($m^3$/month) for many rivers in the world. The record for Amazon River was observed in the downstream Amazon River at a station in Obidos, Brazil from December 1927 to October 2018, totally 1091 months available. The record for Congo River was measured at a station in Kinshasa, Congo from January 1903 to January 2011, totally 1296 months. We calculated moving mean river flow using a moving window of length 3 months and used it as the smoothed river flow for the third month. We took the smoothing approach the reduction in noise resulted in more robust predictions across all models.

For both predictor (SST) and predictand (river flow) our monthly data span from January 1950 to December 2005. Of this totally 672 months, we use the first 600 months as our training data, the following 36 months as our validation data to select best parameters for the model, and the last 36 months (January 2003 to December 2005) as the test data. While our dataset is limited in size by the record length, in the future additional data, including discharge data from additional rivers, can be used to bolster the results.

**Neural Network Model**

The CNN used in this paper consists of 4 convolutional layers and 3 fully connected layers. The number of output channels for each convolutional layer is 32, 32, 64 and 64, respectively. They all have stride 1. The filter sizes in the first three layers are 3×3, and for the fourth layer, it is 1×1. All convolutional layers are followed by a ReLU activation and a 2D max pooling layer with size 2×2 and stride 2×2. For the fully connected layers, the number of output feature for each layer is 128, 64 and 1, respectively. The input image size is 80×300×C with different number of channels *C*. For all ESMs as input, C=32. For all reanalysis input, C=3. For mean ESMs or mean reanalysis as input, C=1. The network output is a scalar. We set the training batch size as 64 and use Adam optimizer with initial learning rate 5×10$^{-5}$ and weight decay 1×10$^{-4}$. We use squared loss function and the network tries to minimize the loss function: $\frac{1}{T}\sum_{t=1}^{T}(f(X_t, w) - y_t)^2$, where *T* is the number of training samples, $X_t \in R^{W \times H \times C}$ is the *t*-th input with width *W*, height *H* and number of channels *C*, y$_t$ is the *t*-th ground truth target, w={w$_1$,…,w$_L$} is the set of weights from all layers. The network output $f(X_t, w) = f_L(f_{L-1}(... f_1(X_t, w_1)))$, where $f_l(., w_l)$ is the mapping function for the *l*-th layer in the neural network. Predictive uncertainty was estimated as the standard deviation of five repeated CNN predictions with different learning rates.

**Saliency Map and Cyclical Saliency Map (Cyclic-SM)**

The saliency map for a CNN is the derivative of the network output *y* with respective to the input $X$: $S = \frac{\partial y}{\partial X} = \frac{\partial f(X,w)}{\partial X}$, where *S* is the same size as the input[15]. The magnitude of elements *S$_{ijk}$* in *S* reflects how important the corresponding input pixel *X$_{ijk}$* (where *i,j,k* is the index of the width, height and channel of *X*) is to the output prediction. For climate variables viewed as images in different time frame, they usually exhibit some (irregular) periodicity in the time. We can utilize this property to enhance the saliency map by superimposing individual saliency maps to form a conglomerate saliency map. Specifically, we define the Cyclic-SM with a cycle *M* as: $S^c = \frac{1}{K+1}\sum_{k=0}^{K} S_{t+kM} = \frac{1}{K+1}\sum_{k=0}^{K} \frac{\partial y_{t+kM}}{\partial X_{t+kM}}$, where $K = \left\lfloor \frac{T-t}{M} \right\rfloor$ is the number of individual saliency maps in the cycle.

The averaging nature of the Cyclic-SM makes it more robust to gradient fluctuation and noise compared to an ordinary saliency map. In addition, Cyclic-SMs are meaningful in climate context. For example, for monthly data, M=12 corresponds to a natural month cycle (January, February, … , December). And we further define seasonal and yearly Cyclic-SM as the sum of saliency maps of the corresponding months. We can calculate different Cyclic-SMs with different cycles depending on the specific purpose and climate data used. For example, we can get daily, monthly, seasonal, annual or other Cyclic-SMs to analyze the dependencies between climate variables in different time scales.

**Data availability**

All data used are publicly available. The ESM data used in this study are available from the NASA Earth Exchange. The SST data (Hadley-OI, COBE, and NOAA ERSSTV5) used in this study are available from UCAR Climate Data Guide. The river flow dataset was obtained from UCAR and can be downloaded from UCAR Research Data Archive.

**Code availability**

Codes are available online at https://github.com/yuminliu/SaliencyMap[39].

**Acknowledgements**

This work was funded by the NSF grants CyberSEES CCF-1442728 (JGD, ARG), Big Data IIS-1447587 (ARG), and SES-1735505 (ARG). KD and ARG acknowledge support from the NASA Ames Research Center.

**Author contributions**

YL performed the analysis and wrote the first draft of the paper. KD helped define the problem and co-wrote the paper with YL and ARG. ARG and JD defined the problem. YL, KD, ARG, and JD interpreted the results and contributed to the writeup.

**Competing interests**

The authors declare no competing interests.

**Materials & Correspondence**

Correspondence to Auroop R. Ganguly, auroop@gmail.com

# Figures

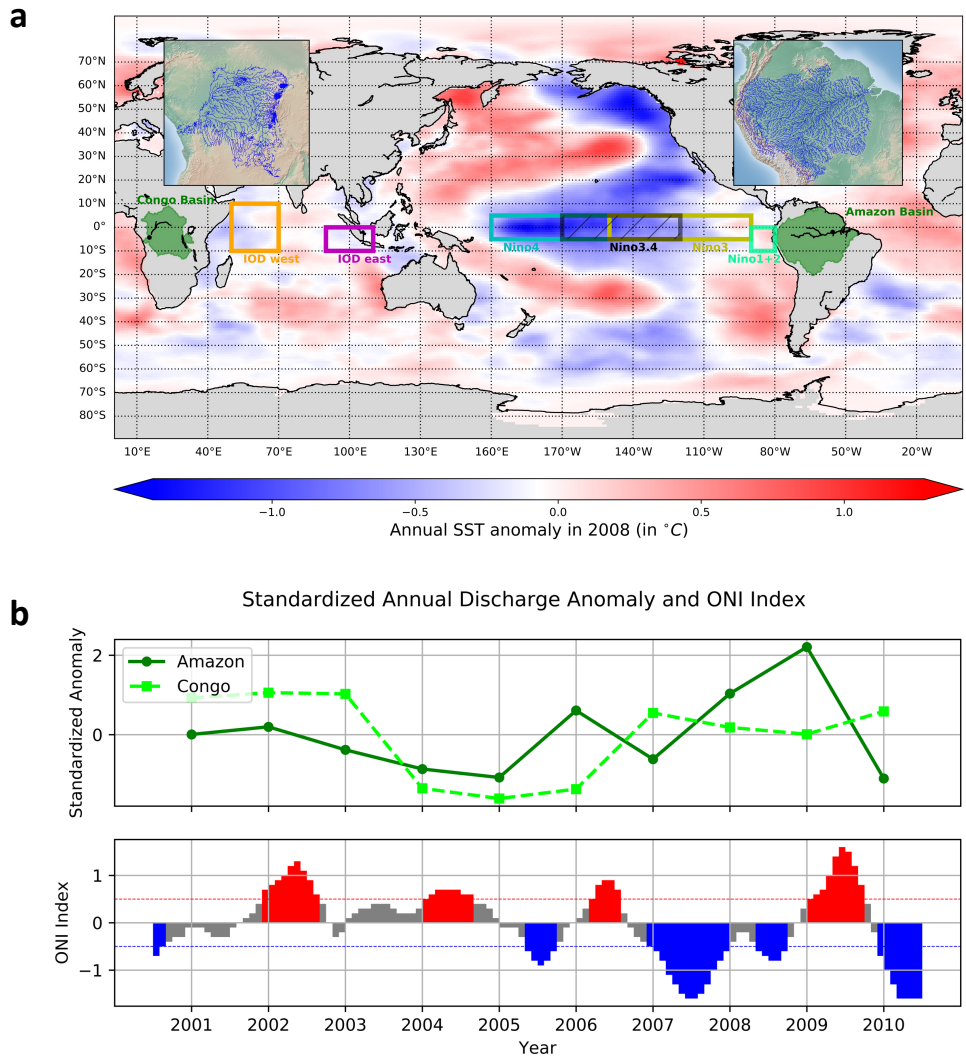

**Figure 1. Global sea surface temperature fluctuations including the El Niño Southern Oscillation impact interannual variability in the flow of large rivers such as Amazon and Congo. a**, Regions for calculating El Niño–Southern Oscillation (ENSO) indices (Niño 1+2, Niño 3, Niño 3.4 and Niño 4) and Indian Ocean Dipole Mode Index (DMI), and two hydrological regions (Amazon River basin and Congo River basin). The colors shown on the ocean is the annual sea surface temperature (SST) anomaly in 2008, a La Niña year. **b**, Time series of standardized annual river flow in $m^3/s$ for Amazon (green) and Congo (lime) and monthly Oceanic Niño Index (ONI) in the Niño 3.4 region at the same time-period. The ONI data are from United States Climate Prediction Center (NOAA 2021). Warm (red) and cold (blue) periods show months that are higher than +0.5°C or lower than -0.5°C threshold for minimum of five consecutive months. A warm/cold year is a year when warm/cold anomaly months dominate, and a neutral year is a year that is neither a warm nor a cold year. For Amazon, the river flow decreases during the warm period and increases during the cold period. However, the relations between Congo River flow and ONI are more complicated and not obvious.

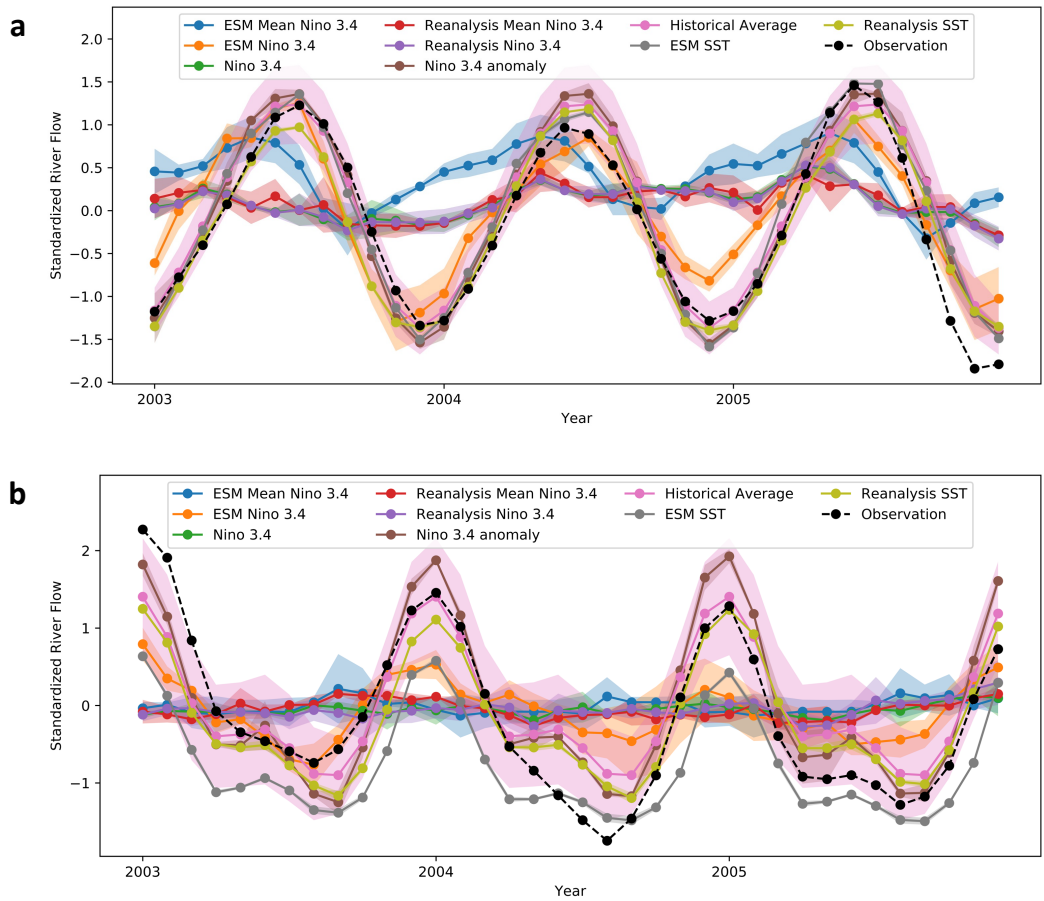

**Figure 2. Predictions of the interannual variability of the Amazon and Congo rivers based on observed and model-simulated sea surface temperatures compared with climatology.** River flow ground truth observations (black) and predictions using different predictors from January 2003 to December 2005 for Amazon (**a**) and Congo (**b**) river. The predictors are mean Niño 3.4 calculated from 32 Earth System Models (ESM) (ESM Mean Niño 3.4), Niño 3.4 calculated from each of 32 ESMs (ESM Niño 3.4), Niño 3.4 index from NOAA (Niño 3.4), Niño 3.4 calculated from 3 reanalysis (Reanalysis Mean Niño 3.4), Niño 3.4 calculated from each of 3 Reanalysis (Reanalysis Niño 3.4), Niño 3.4 anomaly (Niño 3.4 index calculated by NOAA from HadISST1), SST from 32 ESMs (ESM SST, light purple) and SST from 3 reanalysis (Reanalysis SST, gray). Seasonality was subsequently added to the predictions of river flow anomaly based on Niño 3.4 anomaly to generate absolute river flow. The brown line is the historical average prediction result. For models using ENSO index as predictor, we applied six models [linear regression, ridge regression, elastic net regression, random forest regression and deep neural network (DNN) regression] and use their ensemble as the final prediction. The shaded areas are 1 standard deviation for ensemble methods and historical averaging.

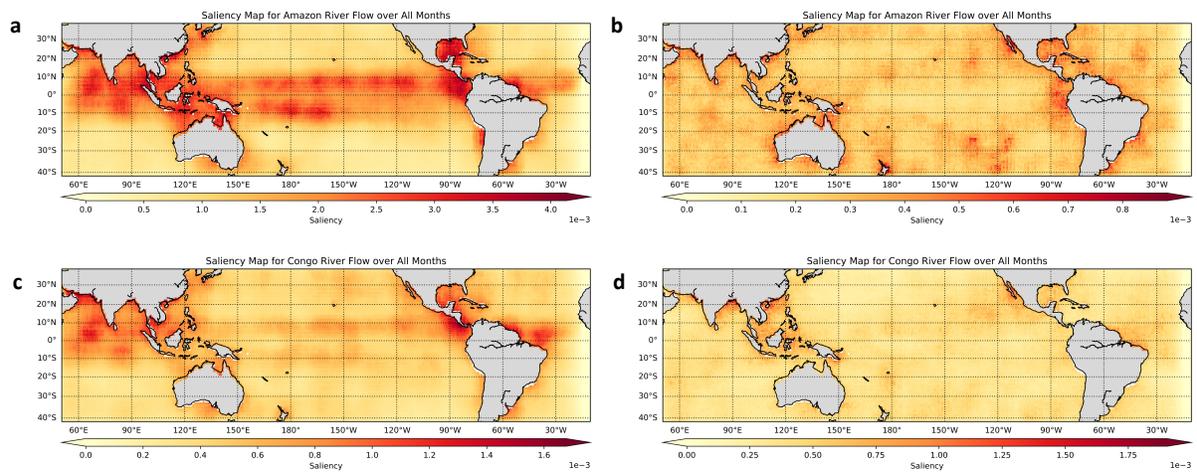

**Figure 3. Explainable deep learning showing saliency maps for predictive understanding with the network model representations. a,b**, Saliency Map for Amazon River flow prediction using ESMs (**a**) and reanalysis (**b**) SST, respectively. **c,d**, Saliency Map for Congo River flow prediction using ESMs (**c**) and reanalysis (**d**) SST, respectively. When using ESM SST as predictor, the salient areas mainly lie in the tropical Pacific and Indian Ocean, but they are much more diffused when using reanalysis SST.

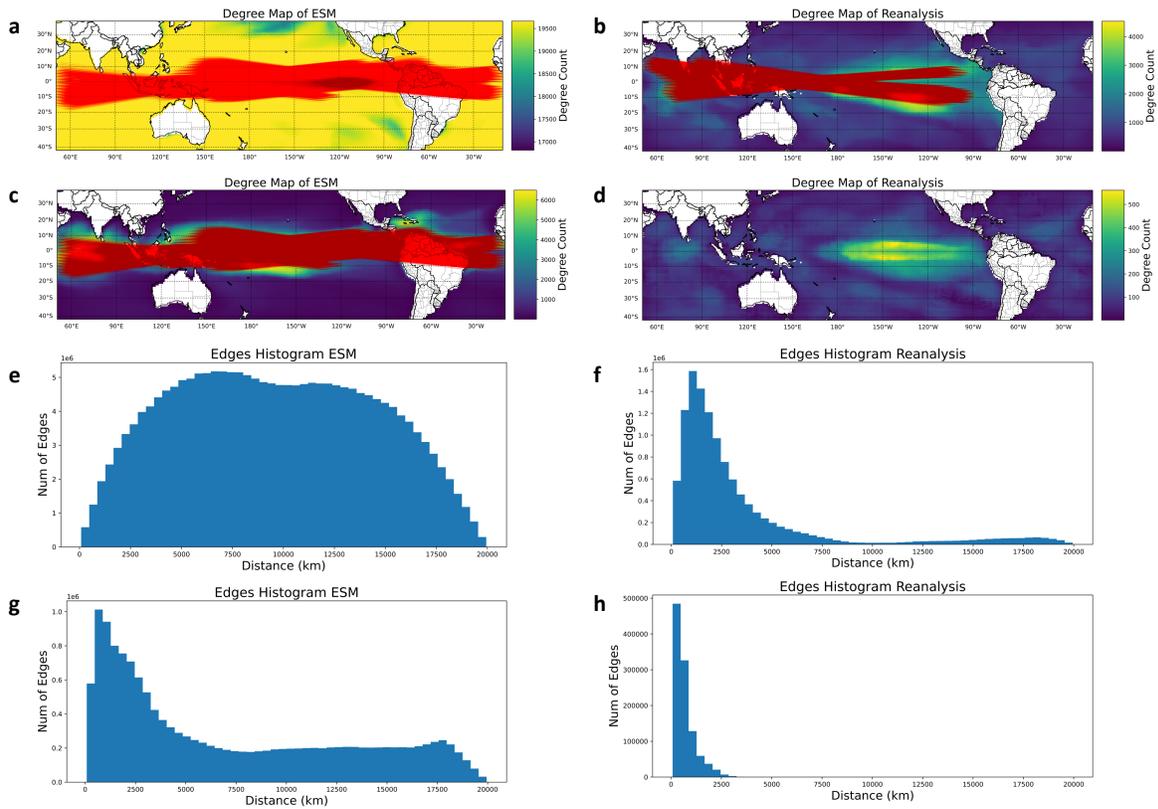

**Figure 4. Teleconnections in space and time based on reanalysis and model-simulated link strengths and degree maps used to construct and interpret complex networks in climate. a,c**, Degree map and teleconnections for mean ESM SSTs. (**a**), correlation threshold equal to 0.5 and 0.9 for degree and teleconnection. **c**, correlation threshold equal to 0.9 and 0.9 for degree and teleconnection. **b,d**, Degree map and teleconnections for mean Reanalysis SSTs. **b**, correlation threshold equal to 0.5 and 0.5 for degree and teleconnection. **d**, correlation threshold equal to 0.9 and 0.9 for degree and teleconnection. We show teleconnections with distance larger than 19000km and 15000km for ESM and Reanalysis SST, respectively. **e,g**, the histogram of edges using correlation threshold 0.5 and 0.9 for mean ESM SST. **f,h**, the histogram of edges using correlation threshold 0.5 and 0.9 for mean Reanalysis SST.

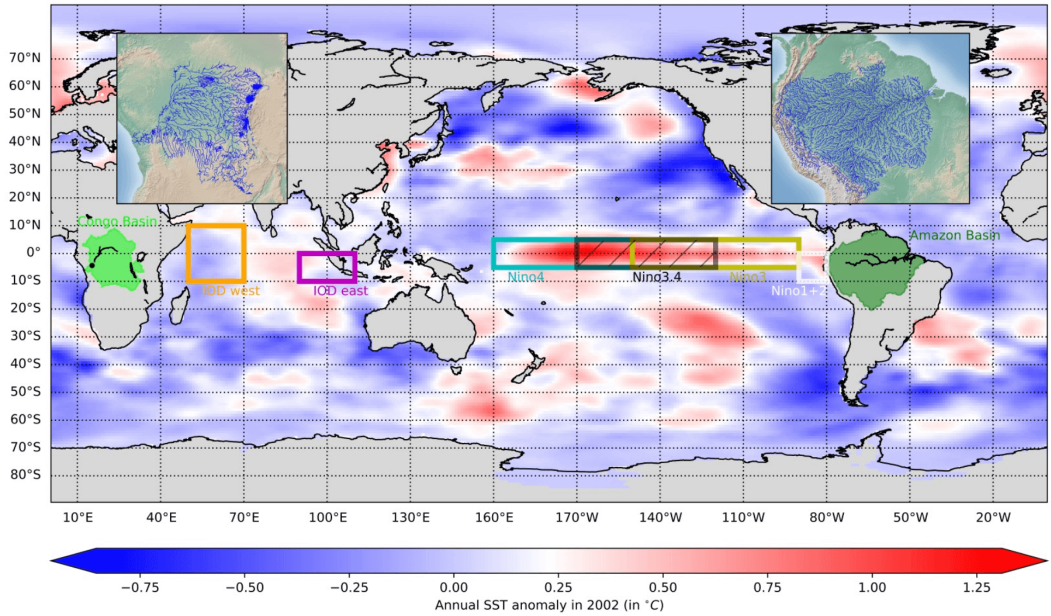

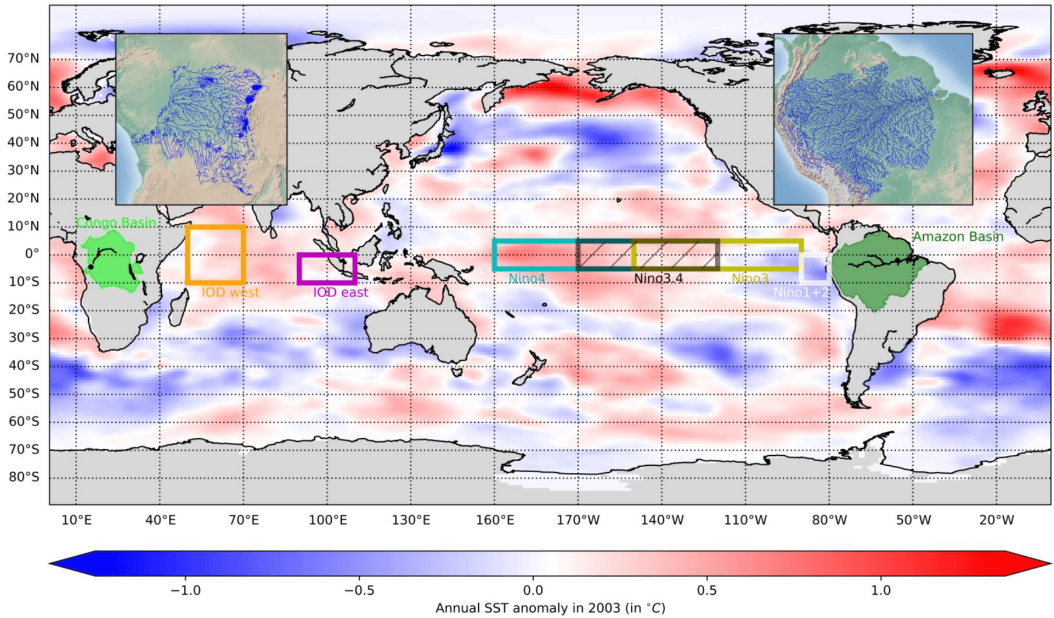

**Figure S1. Global sea surface temperature anomalies in an El Niño year and a neutral year. a**, The year 2002 qualified as an El Niño year because a warm anomaly of +0.5°C or greater persisted for a minimum of 5 consecutive months. Outside of the Niño 3.4 region, the global SST anomalies in 2002 were largely negative. Warming of surface waters the eastern Pacific during El Niño events is associated with weakening of trade winds along the equator, bringing severe rainfall and drought to far-fliung regions. **b**, Neutral states often correspond to transitions between El Niño and La Niña events.

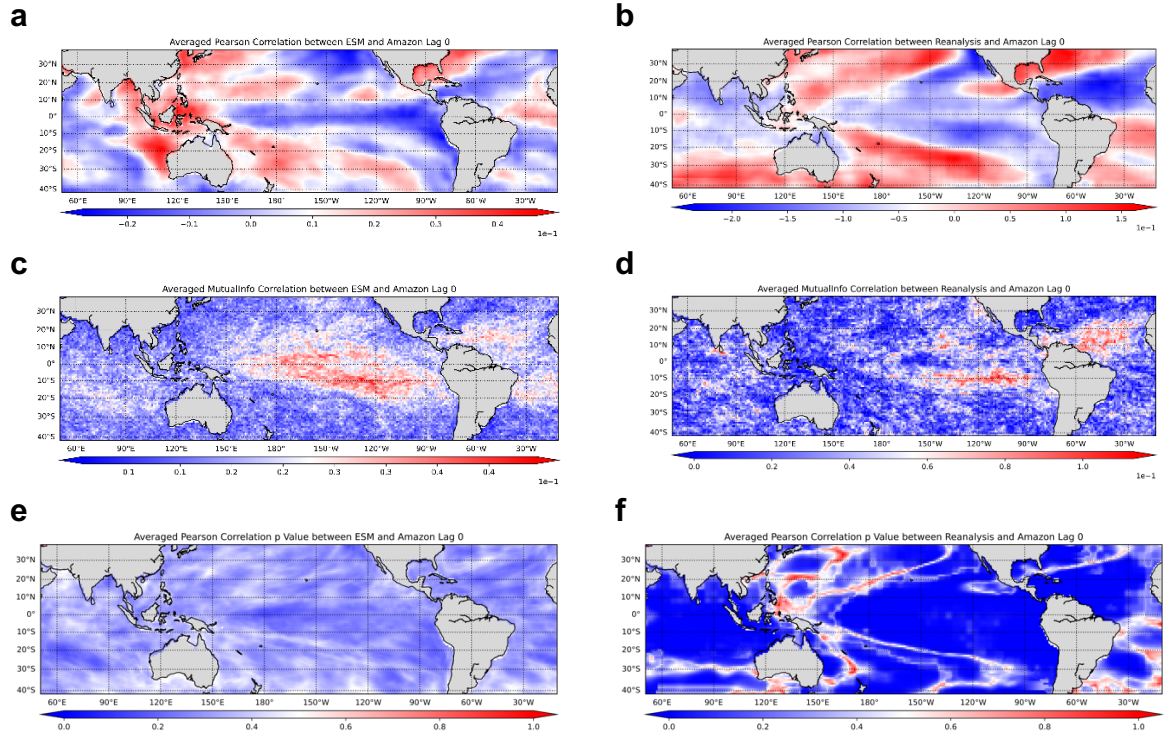

**Figure S2. Linear and nonlinear dependence between Amazon River flow and global sea surface temperature.** Time series of monthly Amazon River flow were analyzed with reference to SST at each geographic location for a period of 672 months from January 1950 to December 2005. Pearson correlation of Amazon River flow with Earth system model (ESM) SST (**a**) and reanalysis SST (**b**) Mutual information between Amazon River flow ESM SST (**c**) and reanalysis SST (**d**). The p-values of Pearson correlation of Amazon River flow with ESM SST (**e**) and reanalysis SST (**f**).

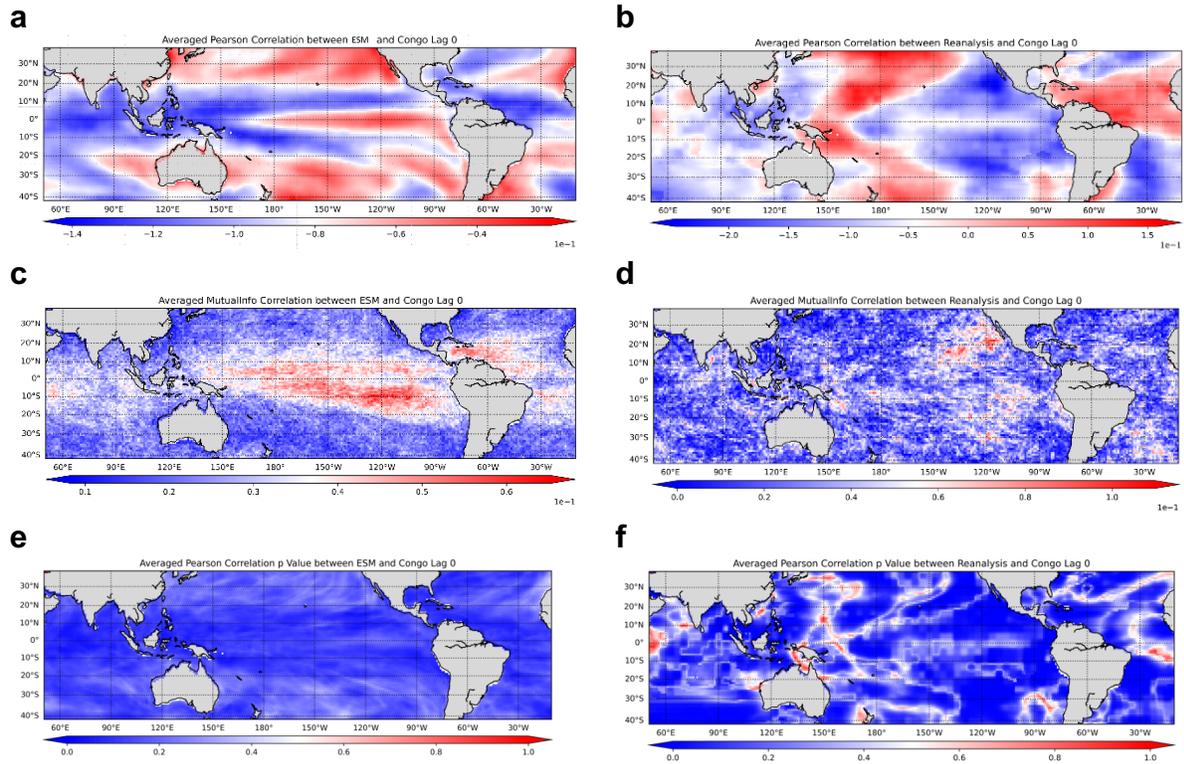

**Figure S3. Linear and nonlinear dependence between Congo River flow and global sea surface temperature.** Time series of monthly Congo River flow were analyzed with reference to SST at each geographic location for a period of 672 months from January 1950 to December 2005. **a,b,** Pearson correlation of Congo River flow with Earth system model (ESM) SST (**a**) and reanalysis SST (**b**) Mutual information between Congo River flow ESM SST (**c**) and reanalysis SST (**d**). The p-values of Pearson correlation of Congo River flow with ESM SST (**e**) and reanalysis SST (**f**).

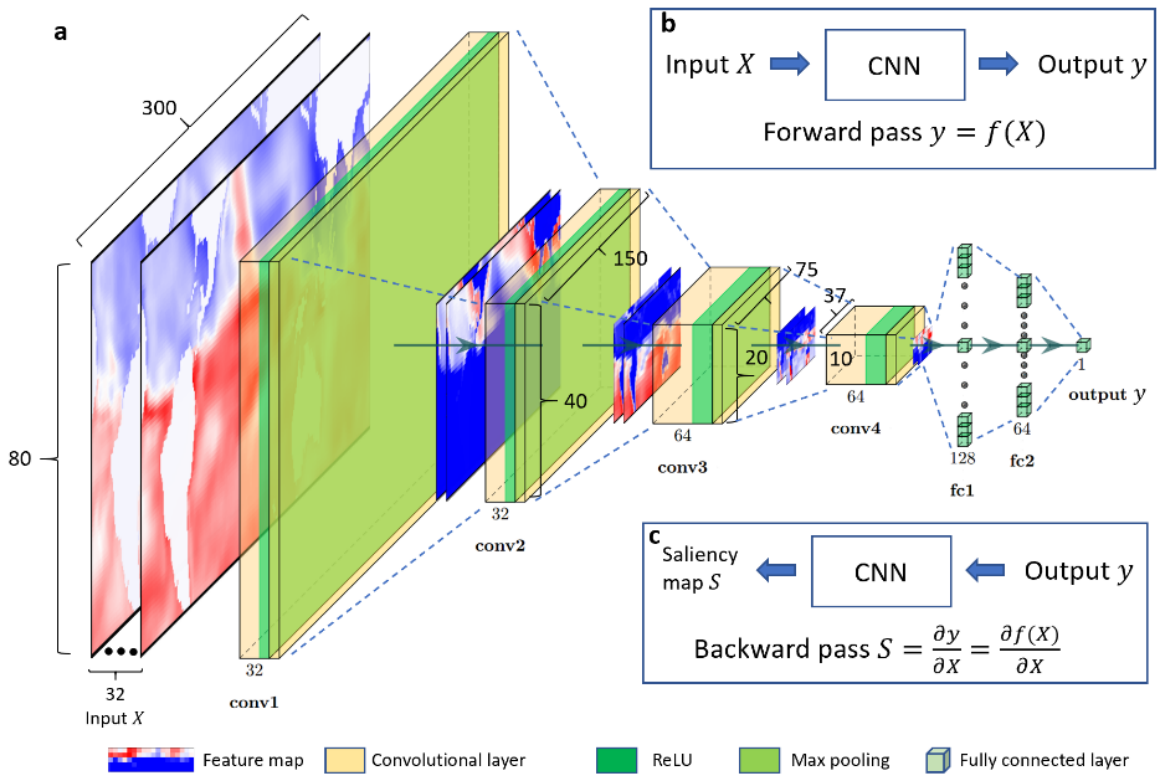

**Figure S4. Architecture of CNN model used in this paper. a**, network architecture. There are 4 convolutional layers (conv1 - conv4) and 3 fully connected layers (fc1,2 and output). Each convolutional layer is followed by a ReLU activation (ReLU($x$)=max(0,$x$)) and a max pooling layer. The image input size is 80×300×$C$ with $C$ = 1, 3 or 32 depending on the datasets. For the convulutional layers, the filter sizes are all 3×3 with stride 1 and padding 1; and the number of filter channel is 32, 32, 64 and 64, respectively. The pooling layers are 2D max pooling layers with size 2×2 and will reduce the feature maps to half size. The outputs of the fully connected layers are 1D vectors with length 128, 64 and 1, respectively. **b**, forward pass to input climate variable SST as $X$ and get prediction $y$. **c**, backward pass to calculate gradient of prediction $y$ with respective to input X to get saliency maps.

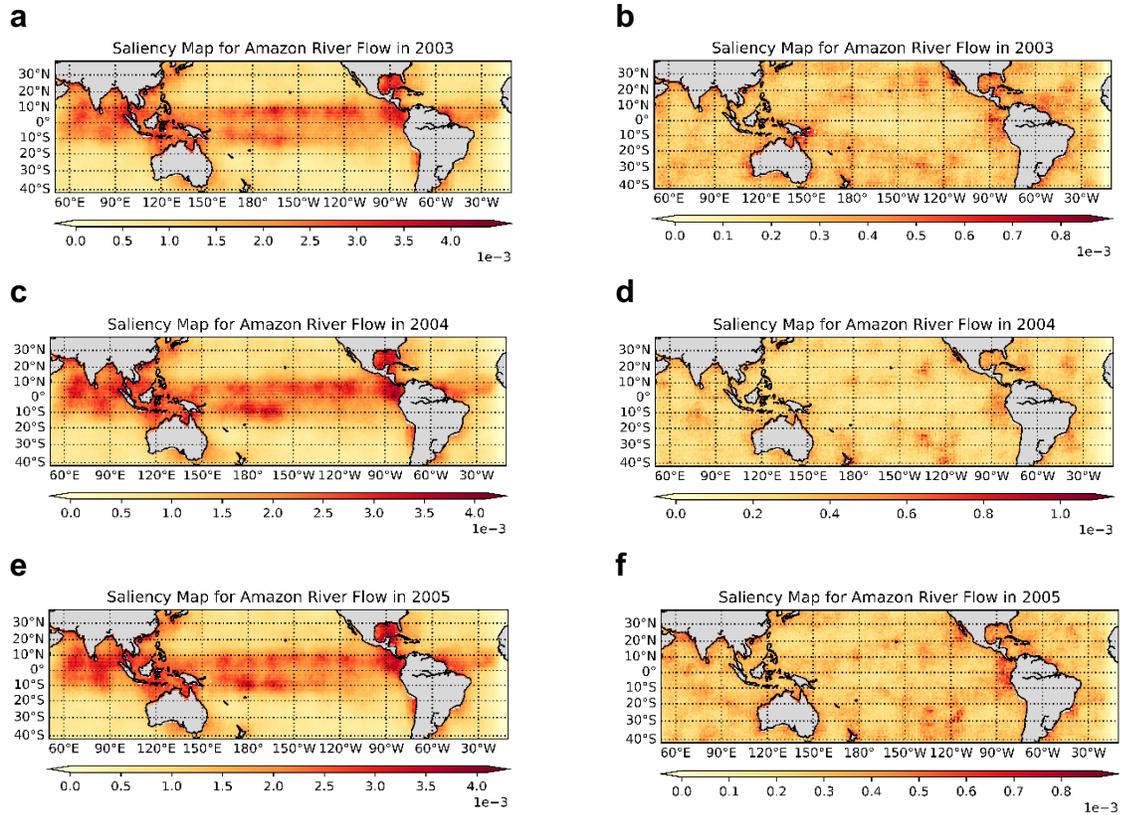

**Figure S5. Yearly cyclical saliency maps highlight the relative contributions of geographic regions to predicting Amazon River flow.** The yearly cyclical saliency maps are calculated as the mean of saliency maps of the 12 months in the corresponding year. **a,c,e** Salient Earth system model (ESM) SST regions for Amazon River flow prediction in the years 2003, 2004, and 2005 were clustered around the equator, but were not limited to the ENSO region. **b,d,f** Salient reanalysis SST regions for Amazon River flow prediction were more diffused across latitudes than ESM regions.

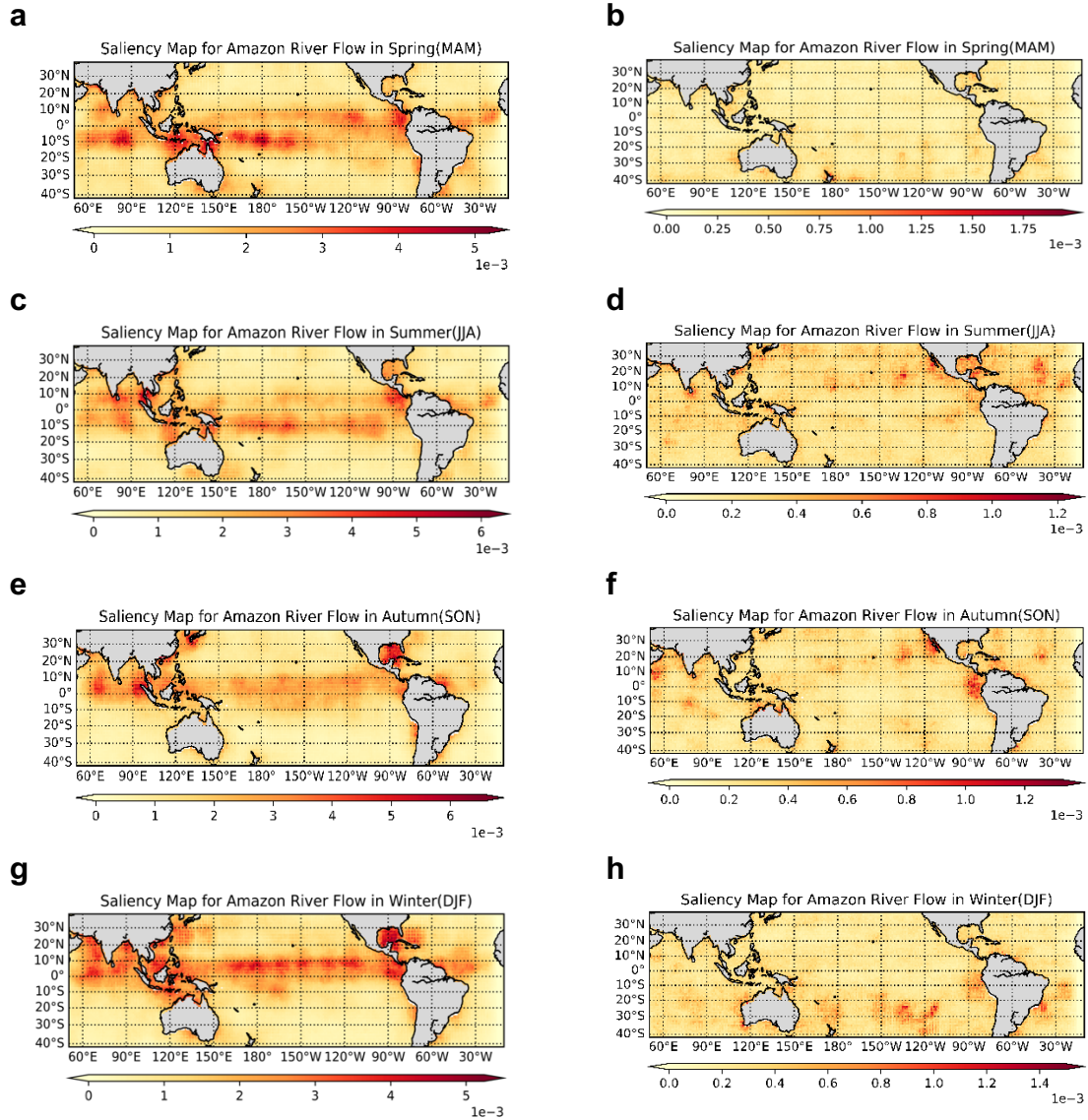

**Figure S6. Seasonal cyclical saliency maps for show periodic changes in the relevant regions for predicting Amazon River flow.** The seasonal cyclical saliency maps are calculated as the mean of saliency maps for different seasons in the Northern Hemisphere (spring: March, April, May; summer: June, July, August; autumn: September, October, November; winter: December, January and February). **a,c,e,g** Saliency maps based on Earth system model SST. **b,d,f,h** Saliency maps based on reanalysis SST suggest more salient regions in the northern latitudes during Northern Hemisphere summer (June, July, August) and in the southern latitudes during Southern Hemisphere summer (December, January, February).

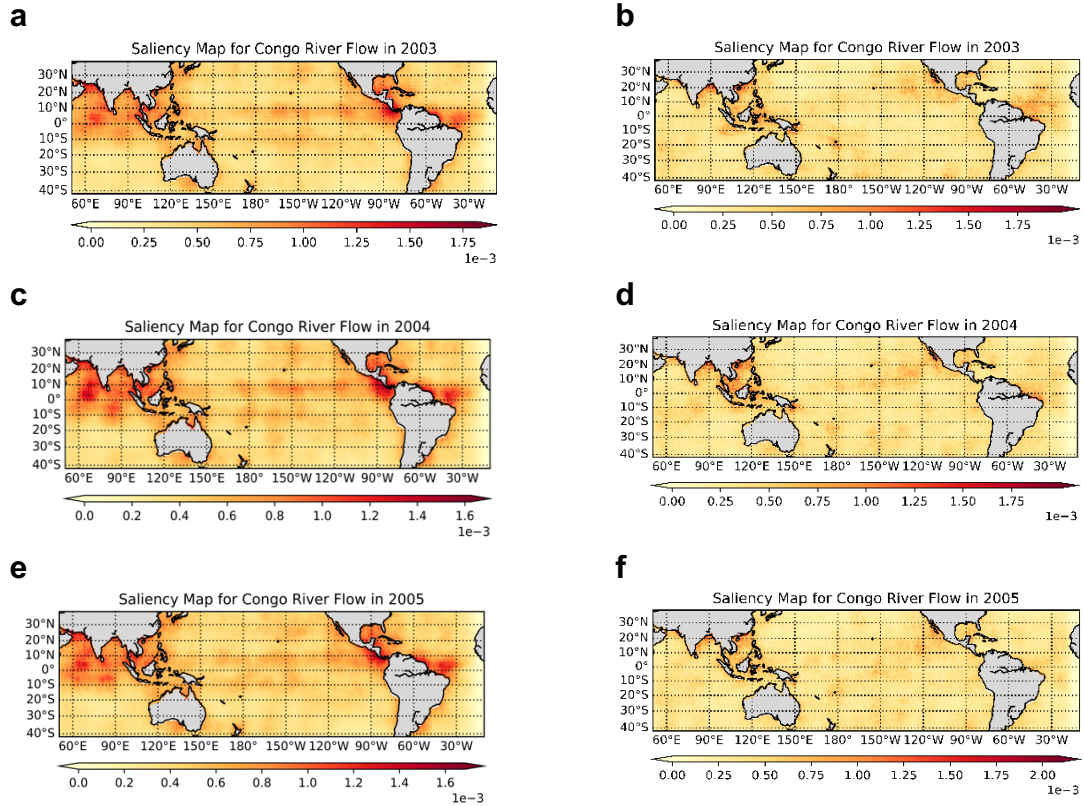

**Figure S7. Yearly cyclical saliency maps highlight the relative contributions of geographic regions to predicting Congo River flow.** The yearly cyclical saliency maps are calculated as the mean of saliency maps of the 12 months in the corresponding year. **a,c,e** Salient Earth system model SST regions for Congo River flow prediction in the years 2003, 2004, and 2005 were clustered around the equator, with Atlantic and Indian ocean regions more salient than the ENSO region. **b,d,f** Salient reanalysis SST regions for Congo River flow prediction were more diffused.

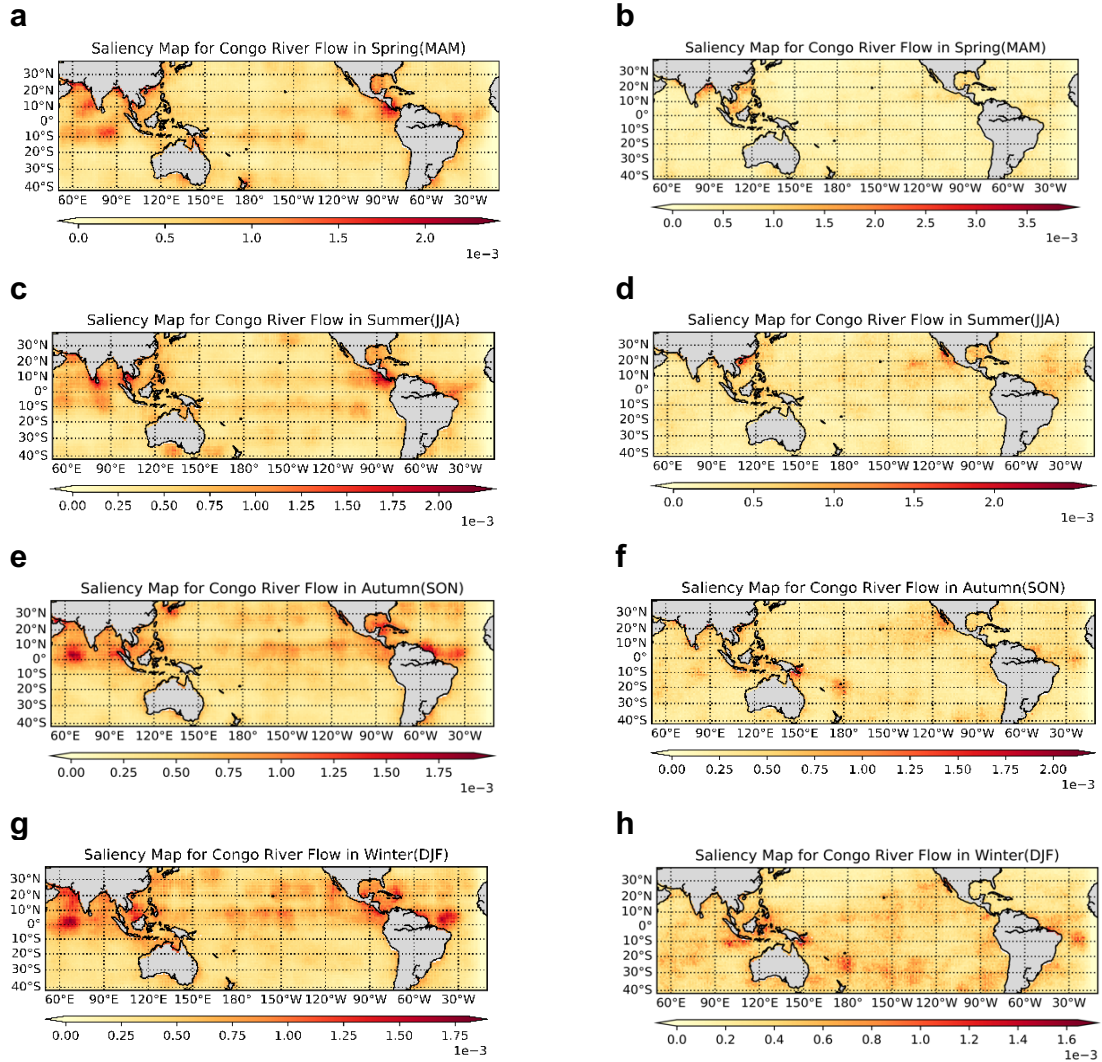

**Figure S8. Seasonal cyclical saliency maps highlight the relative contributions of geographic regions to predicting Congo River flow.** The seasonal cyclical saliency maps are calculated as the mean of saliency maps for different seasons. **a,c,e,g** Saliency maps based on Earth system model (ESM) SST reveal less salient information in the ENSO region than saliency maps for Amazon River prediction. **b,d,f,h** Saliency maps based on reanalysis SST reflect more diffused salient regions than ESM SST.

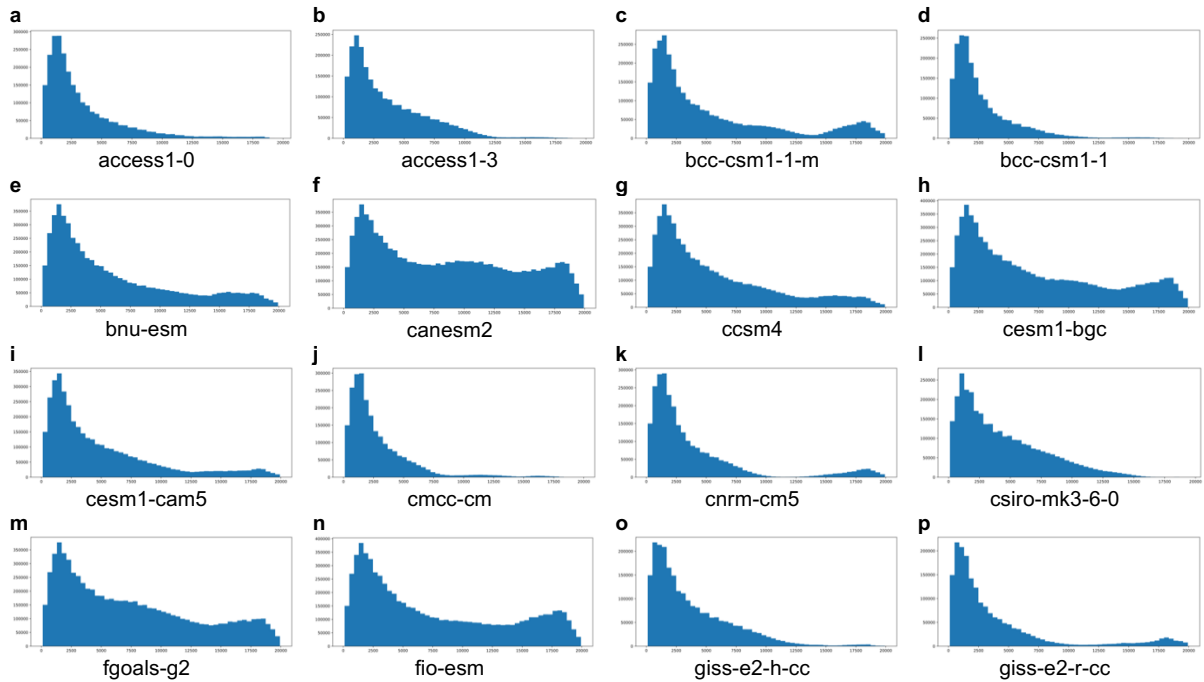

**Figure S9. Diverse climate network topology is suggested by distributions of geographic distance between correlated SST time series.** Defining a connection as any pair of geographic locations where the Pearson correlation between the SST time series is equal or greater than 0.5, the distribution of connection lengths are plotted for each Earth system model (ESM). Variation in the shape of distributions indicates dissimilar topology of climate networks among different ESMs, with substantial variation in the number of long range, or teleconnections. SST time series of 672 months spanning January 1950 to December 2005 were evaluated for locations in the area with latitude from 9.5°S to 9.5°N and longitude from 50.5°E to 349.5°E.

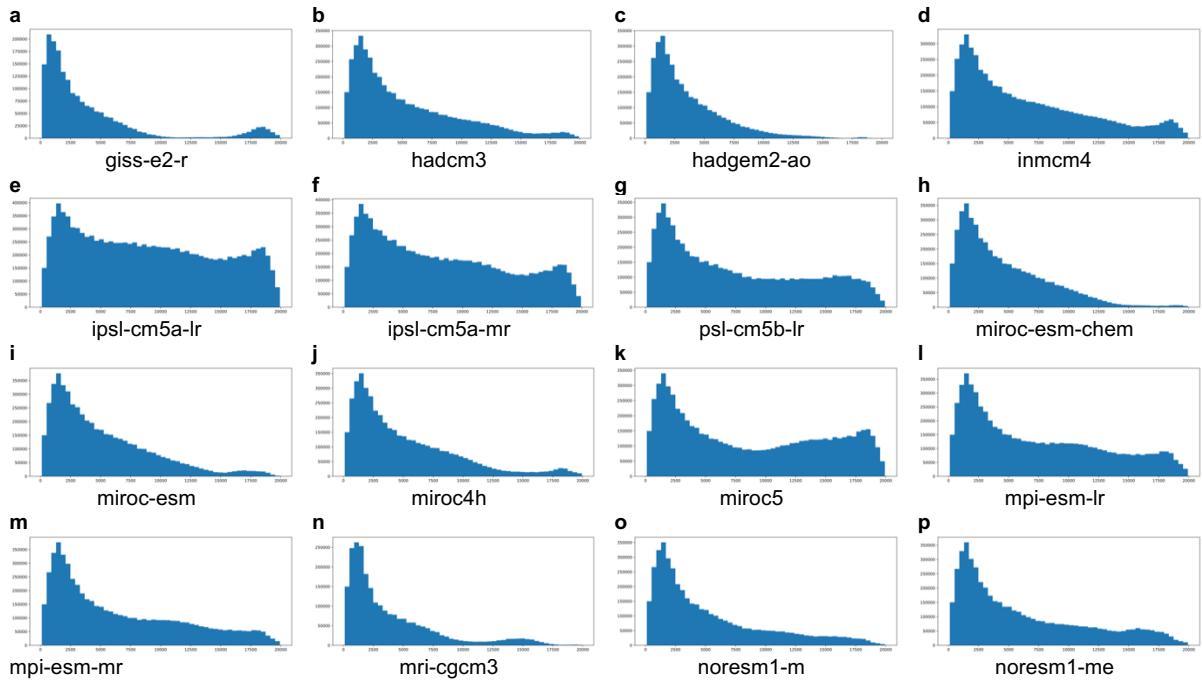

**Figure S10. Diverse climate network topology is suggested by distributions of geographic distance between correlated SST time series (continued).** Defining a connection as any pair of geographic locations where the Pearson correlation between the SST time series is equal or greater than 0.5, the distribution of connection lengths are plotted for each Earth system model.

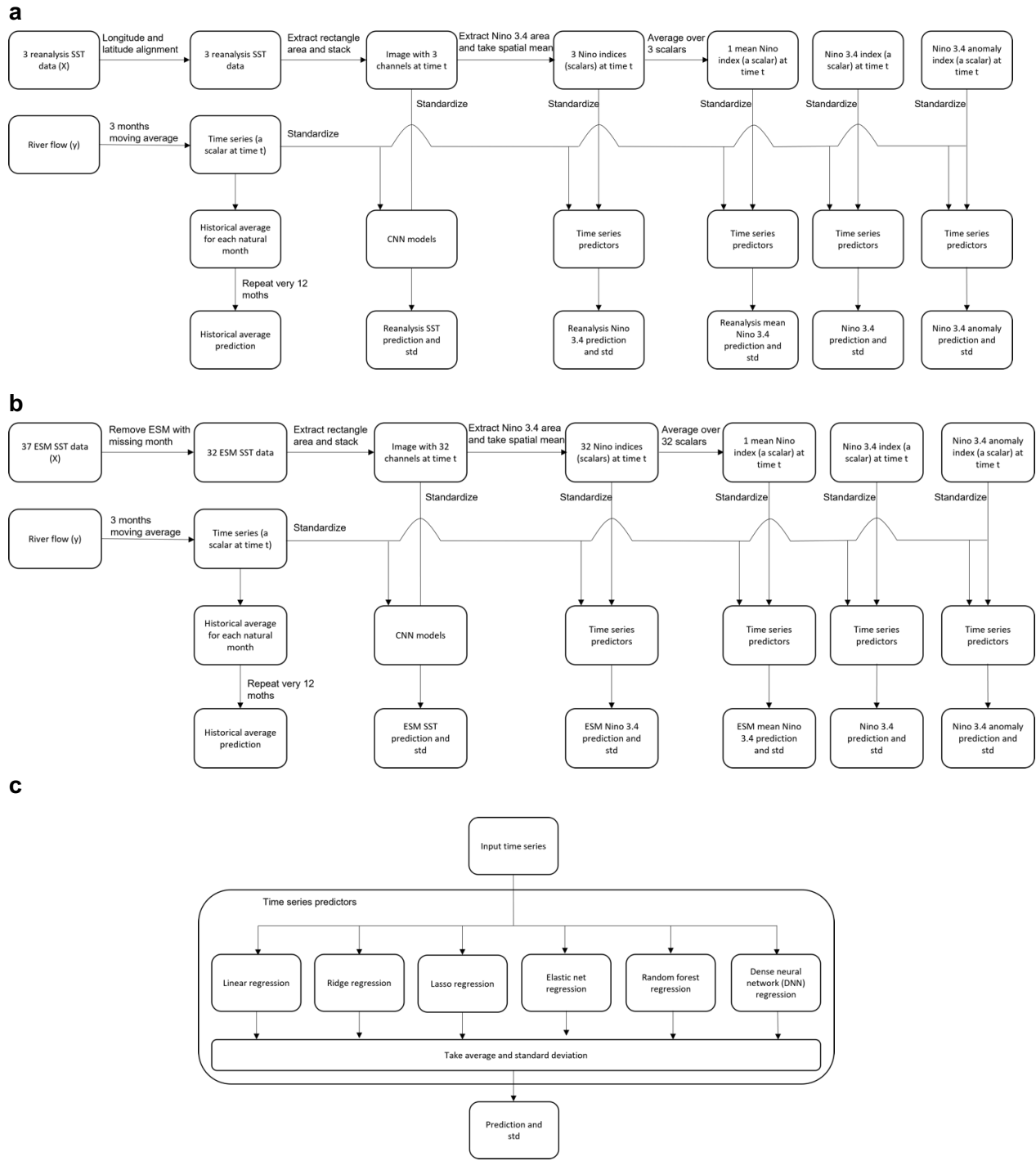

**Figure S11. Flowcharts of dataset processing, river flow modeling, and evaluation of predictions.**
**a,** Reanalysis sea surface temperature is extracted at the Niño 3.4 region and subjected to various levels of spatial aggregation and averaging for use as inputs to ML models. Standardized river flow is used to generate a baseline prediction used on climatological mean, as well as a label for the ML models. **b**, Similar steps are applied to SST simulations from 37 Earth system models. **c**, The ensembling method to generate probabilistic river flow predictions uses a suite of methods to produce a predictions, from which a mean value and predictive variance are calculated.

**Table S1. Earth system models used in the experiments.** All models were obtained from the Coupled Model Intercomparison Project Phase 5.

| Index | Name | Index | Name | Index | Name | Index | Name |
|---|---|---|---|---|---|---|---|
| 0 | access1-0 | 8 | cesm1-cam5 | 16 | giss-e2-r | 24 | miroc-esm |
| 1 | access1-3 | 9 | cmcc-cm | 17 | hadcm3 | 25 | miroc4h |
| 2 | bcc-csm1-1-m | 10 | cnrm-cm5 | 18 | hadgem2-ao | 26 | miroc5 |
| 3 | bcc-csm1-1 | 11 | csiro-mk3-6-0 | 19 | inmcm4 | 27 | mpi-esm-lr |
| 4 | bnu-esm | 12 | fgoals-g2 | 20 | ipsl-cm5a-lr | 28 | mpi-esm-mr |
| 5 | canesm2 | 13 | fio-esm | 21 | ipsl-cm5a-mr | 29 | mri-cgcm3 |
| 6 | ccsm4 | 14 | giss-e2-h-cc | 22 | ipsl-cm5b-lr | 30 | noresm1-m |
| 7 | cesm1-bgc | 15 | giss-e2-r-cc | 23 | miroc-esm-chem | 31 | noresm1-me |

**Table S2. RMSE for predicting Amazon and Congo River flow using Niño 3.4 region SST and larger area (5°S-5°N, 170°W-120°W) SST.** Values with red bold and bold font indicate the best and second-best results on each task, respectively. Values marked with an ∗ indicate the best results when using Niño 3.4 region SST as the predictor.

| | Method | Linear, lasso, ridge, elastic net, random forest and DNN regression | | | | | Niño 3.4 index | Historical average | CNN | |
|---|---|---|---|---|---|---|---|---|---|---|
| | Predictor type | Niño 3.4 average SST (C-dimensional time series, C=1, 3 or 32) | | | | | Niño 3.4 index | Climatological | SST (2-dimensional images) | |
| | Predictor source | ESM mean | ESM | HadISST1 | Reanalysis mean | Reanalysis | HadISST1 | Historical mean | ESM | Reanalysis |
| Amazon | Linear | 1.051 | 0.508* | 0.925 | 0.919 | 0.925 | 0.763 | **0.294** | **0.287** | 0.301 |
| | Ridge | 1.047 | 0.499* | 0.925 | 0.920 | 0.925 | 0.763 | | | |
| | Lasso | 1.050 | 0.576* | 0.925 | 0.919 | 0.925 | 0.763 | | | |
| | Elastic net | 1.028 | 0.469* | 0.925 | 0.922 | 0.925 | 0.762 | | | |
| | Random forest | 1.216 | 0.518* | 0.958 | 1.002 | 0.943 | 0.670 | | | |
| | DNN | 1.002 | 0.516* | 0.954 | 0.929 | 0.954 | 0.770 | | | |
| | Ensemble | 1.049 | 0.461* | 0.931 | 0.925 | 0.928 | 0.950 | | | |
| Congo | Linear | 1.012 | 0.802 | 0.996* | 0.976 | 0.987 | 0.722* | 0.476 | 0.779 | 0.462 |
| | Ridge | 1.012 | 0.784 | 0.996* | 0.977 | 0.985 | 0.722* | | | |
| | Lasso | 1.022 | 0.848 | 0.996 | 0.976 | 1.000 | 0.723* | | | |
| | Elastic net | 1.022 | 0.799 | 0.996 | 0.976 | 0.993 | 0.723* | | | |
| | Random forest | 1.240 | 0.804* | 1.074 | 1.089 | 1.108 | 0.816 | | | |
| | DNN | 1.025 | 0.711* | 1.000 | 0.980 | 0.980 | 0.728 | | | |
| | Ensemble | 1.043 | 0.750 | 1.005 | 0.992 | 0.999 | 1.027 | | | |

**Table S3. Dependence between Niño indices and river flows.** The relationships between river flows and SST are analyzed using Pearson correlation, a measure of linear dependence, and mutual information, a nonlinear measure of the information content shared between two random variables. For the Niño 3.4 region SST, ESM mean is the average SST of 31 Earth system models, HadISST1 is a single reanalysis model, and Reanalysis mean is the average SST of 3 reanalysis models.

| Index | | ESM mean | HadISST1 | Reanalysis mean |
|---|---|---|---|---|
| Amazon | Pearson correlation | -0.0561 | -0.191 | -0.146 |
| | Mutual information | 0.049 | 0.077 | 0.077 |
| Congo | Pearson correlation | -0.195 | -0.024 | -0.091 |
| | Mutual information | 0.058 | 0.008 | 0.097 |

**Table S4. Different metrics for three prediction results for Amazon River.** For Pearson, Spearman and Kendall's Tau correlation, the values in the parenthesis is the correlation and p-value, respectively. For the seasonal RMSE, the values in the parenthesis is RMSE for spring (MAM), summer ( JJA), autumn (SON) and winter (DJF), respectively. For the yearly RMSE, the values in the parenthesis is RMSE for the year 2003, 2004 and 2005, respectively. For the extreme RMSE, the values in the parenthesis is RMSE for predictions whose absolution values are within and outside 2 standard deviations, respectively. For above/below RMSE, the values in the parenthesis is RMSE for predictions whose values are above and below the mean (0), respectively. For ENSO RMSE, the values in the parenthesis is RMSE for warm, cool and neutral months, respectively.

| Metric / Method | Historical average | ESM+CNN | Reanalysis+CNN |
|---|---|---|---|
| Pearson correlation | (0.9637, 4.426e-21) | (0.967, 9.217e-22) | (0.9451, 4.397e-18) |
| Spearman correlation | (0.9504, 8.029e-19) | (0.9681, 5.171e-22) | (0.9284, 3.475e-16) |
| Kendall's tau correlation | (0.8402, 2.067e-12) | (0.8635, 1.265e-13) | (0.7778, 2.485e-11) |
| Mutual information | 1.2425 | 1.2292 | 1.1371 |
| Seasonal RMSE | (0.1695, 0.2220, 0.2407, 0.4576) | (0.2830, 0.2167, 0.1109, 0.3759) | (0.1941, 0.1722, 0.2789, 0.4956) |
| Yearly RMSE | (0.1728, 0.2338, 0.4177) | (0.2904, 0.2362, 0.2652) | (0.2382, 0.1717, 0.4548) |
| Extreme RMSE | (0.2622, 0.3145) | (0.2773, 0.2545) | (0.3339, 0.2998) |
| Above/below RMSE | (0.2815, 0.3057) | (0.2058, 0.3083) | (0.2961, 0.3281) |
| ENSO RMSE | (0.2040, 0.5991, 0.2868) | (0.2794, 0.2333, 0.2611) | (0.1820, 0.6420, 0.3136) |
| MAE | 0.2215 | 0.2078 | 0.2382 |
| Nash–Sutcliffe coefficient | 0.9053 | 0.9231 | 0.893 |

**Table S5**. **Different metrics for three prediction results for Congo River.** As in table S4, for Pearson, Spearman and Kendall's Tau correlation, the values in the parenthesis is the correlation and p-value, respectively. For the seasonal RMSE, the values in the parenthesis is RMSE for spring (DJF), summer (MAM), autumn (JJA) and winter (SON), respectively. For the yearly RMSE, the values in the parenthesis is RMSE for the year 2003, 2004 and 2005, respectively. For the extreme RMSE, the values in the parenthesis is RMSE for predictions whose absolution values are within and outside 2 standard deviations, respectively. For above/below RMSE, the values in the parenthesis is RMSE for predictions whose values are above and below the mean (0), respectively. For ENSO RMSE, the values in the parenthesis is RMSE for warm, cool and neutral months, respectively.

| Metric / Method | Historical average | ESM+CNN | Reanalysis+CNN |
|---|---|---|---|
| Pearson correlation | (0.909, 1.792e-14) | (0.8726, 4.051e-12) | (0.8994, 9.031e-14) |
| Spearman correlation | (0.8544, 3.397e-11) | (0.8682, 6.962e-12) | (0.86, 1.826e-11) |
| Kendall's tau correlation | (0.6866, 9.2312e-09) | (0.6921, 2.873e-09) | (0.673, 7.683e-09) |
| Mutual information | 0.7301 | 0.5913 | 0.7445 |
| Seasonal RMSE | (0.4908, 0.4468, 0.5885, 0.3440) | (0.9007, 0.4162, 0.3427, 0.5716) | (0.5257, 0.4639, 0.5788, 0.3400) |
| Yearly RMSE | (0.4866, 0.5178, 0.4171) | (0.8148, 0.4889, 0.4105) | (0.5434, 0.4873, 0.4172) |
| Extreme RMSE | (0.4122, 0.4874) | (0.4217, 0.6528) | (0.4417, 0.4971) |
| Above/below RMSE | (0.4612, 0.4858) | (0.8170, 0.4494) | (0.4985, 0.4758) |
| ENSO RMSE | (0.6421, 0.3819, 0.3943) | (0.7951, 0.5438, 0.4976) | (0.6423, 0.2926, 0.4176) |
| MAE | 0.3879 | 0.4765 | 0.3895 |
| Nash–Sutcliffe coefficient | 0.7783 | 0.65 | 0.7691 |